\documentclass[twocolumn,secnumarabic,amssymb,superscriptaddress, nobibnotes, nofootinbib,aps, prd]{revtex4-1}

\usepackage{mathrsfs}
\usepackage{physics}
\usepackage{tikz}
\usepackage[caption=false]{subfig}
\usepackage[colorlinks=true,linkcolor=blue,citecolor=blue,urlcolor=blue]{hyperref}
\usepackage{cleveref}
\usepackage{amsmath,amssymb,amsfonts,amsthm}
\usepackage{graphicx}
\usepackage{ragged2e}
\usepackage{float}
\usepackage{bm}

\def\rar{\rightarrow}

\def\p{\partial}
\def\ra{\rangle}
\def\la{\langle}

\def\no{\nonumber}
\def\bea{\begin{eqnarray}}
\def\eea{\end{eqnarray}}
\def\be{\begin{equation}}
\def\ee{\end{equation}}

\def\p{\partial}
\usepackage{filecontents}

\begin{document}

\title{Coherence revival under the Unruh effect and its metrological advantage}%
\author{Jiafan Wang}
\affiliation{School of Physics, Xi'an Jiaotong University, Xi'an, Shaanxi 710049, China}

\author{Jing-Jun Zhang}
\affiliation{School of Physics, Xi'an Jiaotong University, Xi'an, Shaanxi 710049, China}
\author{Jun Feng}%
\email{Corresponding author:\\
j.feng@xjtu.edu.cn}
\affiliation{School of Physics, Xi'an Jiaotong University, Xi'an, Shaanxi 710049, China}
\affiliation{Institute of Theoretical Physics, Xi'an Jiaotong University, Xi'an, Shaanxi 710049, China}
\affiliation{Hefei National Laboratory, Hefei 230088, China}



%
\date{\today}%

\begin{abstract}

In this paper, we investigate the quantum coherence extraction {between} two accelerating Unruh-DeWitt detectors, coupling to a scalar field in $(3+1)$-dimensional Minkowski spacetime. We find that quantum coherence as a nonclassical correlation can be generated through the Markovian evolution of the {detector} system, just like quantum entanglement. However, with growing Unruh temperature, in contrast to monotonously degrading entanglement, we find that quantum coherence exhibits a striking revival phenomenon. For certain detectors' initial state choices, {the} coherence measure will reduce to zero at first {and} then grow to an asymptotic value. We verify such coherence revival by inspecting its metrological advantage on the quantum Fisher information (QFI) enhancement. Since the maximal QFI {bounds} the accuracy of quantum parameter estimation, we conclude that the extracted coherence can be utilized as a physical resource in quantum metrology. \\

\end{abstract}


\maketitle


\section{Introduction}
\label{1}
Nonclassical correlations have long been regarded as a key physical resource. A notable example is quantum entanglement, which plays a central role in many areas of fundamental physics \cite{Entg1,Entg2}, such as quantum information theory, black hole physics, and cosmology. Other types of quantum correlations beyond entanglement, such as nonlocality \cite{Entg3} and quantum discord \cite{Entg4}, were also exploited. However, there are more general notions on quantum {resources} characterizing the nonclassicality that makes a system inherently quantum. In recent years, intensive works have been done on quantum coherence, which encapsulates the defining features of quantum theory in arbitrary dimensions, from {the} superposition principle of single qubit to quantum correlations among multipartite. Quantum coherence constitutes a powerful resource \cite{Coh1,Coh2,Coh3}, which was wildly applied in quantum metrology \cite{Coh4}, biological systems \cite{Coh5} and thermodynamics \cite{Coh6}. Moreover, {on} a fundamental level, it has been shown \cite{Coh7,Coh8} that coherence and entanglement are intimately related as they are interchangeable and can be unified into a single resource (see also \cite{Coh9,Coh10} for a review).

On the other hand, it was realized that {a designed physical system can extract the intrinsic nonclassical correlations of quantum vacuum}. For instance, two initially uncorrelated Unruh-DeWitt (UDW) detectors in Minkowski spacetime can generate nonlocal entanglement under relativistic acceleration motion \cite{Entg6,Entg7}, and be witnessed by various measures, such as uncertainty relation \cite{F1,F2} and Bell-type inequalities \cite{Entg7+,F3}. In this context, the residual entanglement of {the} final equilibrium is a result of competition between the Unruh decoherence and entanglement generation from {the} Markovian evolution of the detectors that interacts {with the} background field. In other words, even with the entanglement generation, {the} Unruh effect solely plays the role {of} an environment decoherence \cite{U1}.

Beyond entanglement among multipartite quantum systems, it is intriguing to exploit the possibility of extracting other measures on the quantumness of quantum vacuum. Earlier study \cite{Coh-U1} has observed that the extraction of quantum coherence is possible even for \emph{single} UDW detector. For certain initial energy of the background field and the duration of the interaction, a detector moving with uniform acceleration can extract a larger amount of coherence from field states compared to a detector at rest. In other words, the rate of coherence loss may sometimes become slower for a moving detector than for a detector at rest. Similarly, for a spatially extended UDW detector, coherence harvesting and swelling were also observed \cite{Coh-U2}. Such extracted coherence was shown to be catalytic, meaning the same amount {of} coherence can be repeatedly extracted.

In this paper, we concern with the quantum coherence extraction for a system of \emph{multi}-UDW detectors via a coupling to a massless scalar field in $3+1$ Minkowski spacetime. In particular, we exploit the quantum coherence of a combined system with two UDW detectors, both under a uniform acceleration. Motivated by the entanglement generation for multi-UDW detectors \cite{Entg6}, we anticipate that quantum coherence can also be generated after the system {approaches} its equilibrium, even with two detectors initially in an incoherent state. Starting from general initial states, however, we are interested in the necessary conditions that detectors should satisfy to ensure the coherence extraction, i.e., the residual coherence at equilibrium is larger than the initial quantum coherence of detectors.  

While quantum fluctuation of {the} background field plays a role of an environment, we could treat two-detectors as an open quantum system \cite{Op1}, whose reduced density matrix evolves via a Lindblad-type master equation derived by tracing out the environmental degrees of freedom. For the detectors with uniform acceleration $a$, this formalism may lead us to a naive expectation that just as illustrated for entanglement in various scenarios \cite{Entg8,Entg9}, Unruh effect or environmental decoherence should also result {in} a monotonous decay of quantum interference in two-detectors system. However, we find that quantum coherence indeed exhibits a striking \emph{revival} phenomenon that for {a} certain choice of detectors' initial state, coherence measure \cite{Coh1} will reduce to zero at first then grow up and eventually approach to an asymptotic value, while Unruh temperature $T_U=a/2\pi$ keep growing during the process. 

As an important physical resource, we expect {the} above coherence revival to be verified in some practical quantum process by providing enhanced performance. In metrological tasks, nonclassicality rather than entanglement is necessary to achieve quantum advantages \cite{metro1,metro2}. As a particular nonclassicality which is more general than entanglement, quantum coherence is intimately related to the quantum Fisher information (QFI) \cite{Coh11,Coh12}, a key quantity bounding the ultimate precision of the estimation of a parameter encoded in a quantum state\footnote{For instance, in the context of quantum thermodynamics, coherence cost of any state is determined by its QFI \cite{Coh13}.}. Therefore, motivated by the close relation between the QFI and quantum coherence, we will evaluate the QFI for the two UDW detectors, with Unruh temperature as a parameter chosen to be optimally estimated. We will show that the revival coherence corresponds to the enhanced QFI {and} thus may provide a significant quantum advantage in metrological tasks.

The paper is organized as follows. In Sec. \ref{2}, we review the open quantum system approach to resolve the reduced state of two accelerating UDW detectors. In Sec. \ref{3}, we evaluate the quantum coherence of {the} final equilibrium state of {the} two-detectors system and demonstrate the related coherence revival phenomenon. We obtain the condition {of the} initial state preparation for detectors to ensure coherence revival. In Sec. \ref{4}, we evaluate the QFI of Unruh temperature estimation and show a significant enhancement of QFI can be obtained due to the coherence resource. We conclude in Sec. \ref{5} with some remarks on {the} possible root of coherence revival and its distinction to entanglement. Throughout the paper, we use units where $k_\mathrm{B}=c=\hbar=1$.

\section{Accelerating detectors as an open quantum system}
\label{2}
To proceed, we first recall the dynamics of two accelerating UDW detectors in $3+1$-dimensional Minkowski spacetime \cite{Entg6}. With a coupling to the bath of fluctuating background, we treat two detectors as an open quantum system whose density matrix is governed by a Lindblad master equation and evolves non-unitarily due to environment decoherence or dissipation.

Modeling each detector as a two-level atom, the total Hamiltonian of the combined system of detectors and background field is
\be
H=\frac{\omega}{2}\Sigma_3+H_\Phi+\mu H_I,             \label{eq1}
\ee
where $\omega$ is the energy level spacing of the atom and $\Sigma_3$ is one of symmetrized bipartite operators $\Sigma_i\equiv\sigma_i^{(A)}\otimes\mathbf{1}^{(B)}+\mathbf{1}^{(A)}\otimes\sigma_i^{(B)}$, defined by Pauli matrices $\{\sigma^{(\alpha)}_i|i=1,2,3\}$ with superscript $\{\alpha=A, B\}$ labeling distinct atoms. $H_\Phi$ is the Hamiltonian of free massless scalar fields $\Phi(t,\bm{x})$ satisfying standard Klein-Gordon equation. The interaction Hamiltonian can be written in a dipole form between atoms and fluctuating field bath \cite{Op2}, say $H_I= (\sigma_2^{(A)}\otimes\mathbf{1}^{(B)})\Phi(t,\bm{ x}_A)+(\mathbf{1}^{(A)}\otimes\sigma_2^{(B)})\Phi(t,\bm{ x}_B)$, where $\bm{ x}_A$ and $\bm{ x}_B$ label the space positions of two atoms at same time $t$.

Assuming a weak coupling between two-atom system and environment ($\mu\ll1$), such that the initial state of combined system is approximated as $\rho_{\text{tot}}(0)=\rho_{AB}(0)\otimes |0\ra\la0|$, where $\rho_{AB}(0)$ is the initial state of the detectors and $|0\ra$ is the field vacuum. Obviously, the dynamics of $\rho_{\text{tot}}$ should be governed by a unitary evolution via von Neumann equation $\dot{\rho}_{\text{tot}}(\tau)=-\mathrm{i}[H,\rho_{\text{tot}}(\tau)]$, where $\tau$ is the proper time of the atom. Whenever the typical time scale of the environment is much smaller than that of the detectors \cite{Op1}, we can further assume the detectors undergoing a Markovian evolution\footnote{The reliability of Markovian approximation is a delicate issue. However, for weak coupling, possible non-Markovian correction may be robust only at early-time evolution \cite{nonmarkov}. Since only the asymptotic equilibrium state of the detector is concerned, we can safely neglect the difference between Markovian and non-Markovian solutions.}. The reduced dynamics of the detectors can be obtained by integrating over the background field degrees from the $\rho_{\text{tot}}(\tau)$, driven by a quantum dynamical semigroup of completely positive map. Eventually, the open system dynamics should be governed by a Lindblad-type master equation \cite{Op3,Op4}
\be
\frac{\partial\rho_{AB}(\tau)}{\partial \tau}=-\mathrm{i}[H_{\tiny\mbox{eff}},\rho_{AB}(\tau) ]+\mathcal{L}\left[\rho_{AB}(\tau)\right], 
\label{eq2}
\ee
where
\be
\mathcal{L}\left[\rho\right]=\sum_{\substack{i,j=1,2,3\\   \alpha,\;\beta=A,B}}\frac{C^{(\alpha\beta)}_{ij}}{2}\left[2\sigma_j^{(\beta)}\rho_{AB}\sigma_i^{(\alpha)}-\{\sigma_i^{(\alpha)}\sigma_j^{(\beta)},\rho_{AB}\}\right],       \label{eq3}
\ee
representing a dissipative evolution attributed to the interaction between the detectors and {the} external field. The Kossakowski matrix $C^{(\alpha\beta)}_{ij}$ can be explicitly resolved. After introducing the Wightman function of scalar field $G^{(\alpha\beta)}\left(x, x^{\prime}\right)=\left\langle 0\left|\Phi(t,\bm{x}^{(\alpha)}) \Phi\left({t}^{\prime},\bm{x}^{(\beta)}\right)\right| 0\right\rangle$, its Fourier transform is
\be
\mathcal{G}^{(\alpha\beta)}(\lambda)=\int_{-\infty}^{\infty}\mathrm{d}\tau~e^{\mathrm{i}\lambda\tau}G^{(\alpha\beta)}(\tau),
 \label{eq4}
\ee
where the superscript $\alpha,\beta=\{A,B\}$ labeling distinct atoms. For two-atom system, one can easily find that $G^{(AA)}=G^{(BB)}$ and $G^{(AB)}=G^{(BA)}$, which lead to $\mathcal{G}^{(AA)}=\mathcal{G}^{(BB)}\equiv\mathcal{G}_0$ and $\mathcal{G}^{(AB)}=\mathcal{G}^{(BA)}$.

The master equation (\ref{eq2}) enables us to describe the asymptotic equilibrium states of detectors at large times, which are determined by the competition between environment dissipation and quantum correlations generated through the Markovian evolution of detectors \cite{Entg6}. For two-atom system, the initial interatomic separation $L\equiv|\bm{x}^{(A)}-\bm{x}^{(B)}|$ is a control parameter of correlation generation. Thus, the Kossakowski matrices now become distance-dependent since in general $\mathcal{G}^{(AB)}=\mathcal{G}^{(BA)}\equiv\mathcal{G}(\omega,L)=\mathcal{G}_0(\omega)f(\omega,L)$ for two separated atoms \cite{Op2}, where $f(\omega,L)$ is an even function of frequency $\omega$. One would not be surprised that the correlation generation between atoms would be more effective for smaller $L$, and becomes impossible for an infinitely large separation. In fact, it was shown \cite{Entg8} that there always exists a proper $L$, below which the generated correlations can persist under environment dissipation. Therefore, we can concisely fix a small interatomic separation and only {be concerned} about the influence of environment decoherence on the equilibrium states of detectors. In such a situation, all the Kossakowski matrices become equal $C^{AA}_{ij}=C^{BB}_{ij}=C^{AB}_{ij}=C^{BA}_{ij}\equiv C_{ij}$ \cite{Plus1}, where the coefficients $C_{ij}$ are determined by a decomposition
\be
C_{ij}=\frac{\gamma_+}{2}\delta_{ij}-\mathrm{i}\frac{\gamma_-}{2}\epsilon_{ijk} \delta_{3k}+\gamma_0\delta_{3i}\delta_{3j}   ,       \label{eq5}
\ee
where
\be
\gamma_\pm= \mathcal{G}(\omega)\pm \mathcal{G}(-\omega),~~~\gamma_0=\mathcal{G}(0)-\gamma_+/2 .        \label{eq6}
\ee
Moreover, the interaction with {the} external scalar field would induce a Lamb shift contribution for the detector effective Hamiltonian $H_{\mbox{\tiny eff}}=\frac{1}{2}\tilde{\omega}\sigma_3$, in terms of a renormalized frequency $\tilde{\omega}=\omega+\mathrm{i}[\mathcal{K}(-\omega)-\mathcal{K}(\omega)]$, where $\mathcal{K}(\lambda)=\frac{1}{\mathrm{i}\pi}\mbox{P}\int_{-\infty}^{\infty}\mathrm{d}\omega\frac{\mathcal{G}(\omega)}{\omega-\lambda}$ is Hilbert transform of Wightman functions. 

Following a trajectory of accelerating detectors, one can find that the field Wightman function fulfills the Kubo-Martin-Schwinger (KMS) condition, i.e., $G^{+}(\tau)=G^{+}(\tau+\mathrm{i} \beta)$, where $\beta\equiv1/T_U=2\pi/a$ is recognized. Translating it into frequency space, one has 
\be
\mathcal{G}(\lambda)=e^{\beta\omega}\mathcal{G}(-\lambda).        \label{eq7}
\ee
Using translation invariance $\langle 0|\Phi(x(0)) \Phi(x(\tau))| 0\rangle=\langle 0|\Phi(x(-\tau)) \Phi(x(0))| 0\rangle$, after some algebras, we find (\ref{eq6}) can be resolved as
\bea
\gamma_+&=&\int_{-\infty}^{\infty}\mathrm{d}\tau~e^{\mathrm{i}\lambda\tau}\la0|\left\{\Phi(\tau),\Phi(0)\right\}|0\ra=\left(1+ e^{-\beta\omega}\right) \mathcal{G}(\omega),\no\\
\gamma_-&=&\int_{-\infty}^{\infty}\mathrm{d}\tau~e^{\mathrm{i}\lambda\tau}\la0|\left[\Phi(\tau),\Phi(0)\right]|0\ra=\left(1- e^{-\beta\omega}\right) \mathcal{G}(\omega),\no\\
\label{eq8}
\eea
holding true for generic interacting fields. For later use, we also introduce the ratio
\be
\gamma\equiv\gamma_-/\gamma_+=\frac{1- e^{-\beta\omega}}{1+ e^{-\beta\omega}}=\tanh(\beta\omega / 2) ,       \label{eq9}
\ee
which depends solely on the Unruh temperature $T_U$ due to the frequency KMS condition (\ref{eq7}).

By expressing the reduced density matrix of the two-atom system as
\be
\rho_{AB}(\tau)=\frac{1}{4}\left[\mathbf{1}^A\otimes\mathbf{1}^B+\sum_{i=1}^3\rho_{i}\Sigma_i+\sum_{i,j=1}^3\rho_{ij}\sigma_i^A\otimes\sigma_j^B\right],
\label{eq10}
\ee
and inserting it back into (\ref{eq2}), one can derive \cite{F1} the reduced density matrix of two-UDW detectors at equilibrium in an X-type structure
\be
\rho_{AB}(\omega,\beta,\Delta_0)=\left(\begin{array}{cccc}
A & 0 & 0 & 0 \\
0 &C & D & 0 \\
0 & D & C & 0 \\
0 & 0 & 0 & B 
\end{array}\right),
\label{eq11}
\ee
where
\bea
\displaystyle A&=&\frac{(3+\Delta_0)(\gamma-1)^{2}}{4\left(3+\gamma^{2}\right)}~~~~,~~~~
\displaystyle B=\frac{(3+\Delta_0)(\gamma+1)^{2}}{4\left(3+\gamma^{2}\right)}, \no\\
\displaystyle C&=& \frac{3-\Delta_0-(\Delta_0+1) \gamma^{2}}{4\left(3+\gamma^{2}\right)}~~,~~
\displaystyle D= \frac{\Delta_0-\gamma^{2}}{2\left(3+\gamma^{2}\right)} .   \label{eq12}
\eea

We observe that the final equilibrium state of {the} two-detectors system now depends on the ratio $\gamma$ characterizing the thermal nature of {the} Unruh effect, as well as the choices of initial state encoded in $\Delta_0=\sum_i\text{Tr}[\rho_{AB}(0)\sigma^{(A)}_i\otimes\sigma^{(B)}_i]$, a dimensionless constant of motion satisfying $-3\leqslant\Delta_0\leqslant1$ to keep $\rho_{AB}(0)$ positive.

\section{Quantum coherence and its revival}
\label{3}

\subsection{Quantum coherence monotone}
We first review some elementary concepts concerning coherence measures. The notion of coherence admitted in this paper is identified {as} the one in \cite{Coh1}, where {the following set of axioms should be satisfied to specify a reasonable measure of quantum coherence}.

For a given fixed basis $\{|i\rangle\}$, the set of \emph{incoherent states} $\mathcal{I}$ is the set of quantum states with diagonal density matrices with respect to this basis. Incoherent completely positive and trace-preserving maps (ICPTP) are maps that map every incoherent state to another incoherent state. Given this, we say that $\mathcal{C}$ is a proper measure of quantum coherence if it satisfies {the} following properties: 

(C1). $\mathcal{C}(\rho) \geqslant 0$ for any quantum state $\rho$ and equality holds iff $\rho \in \mathcal{I}$. 

(C2a). The measure is non-increasing under a ICPTP map $\Phi$, i.e., $\mathcal{C}(\rho) \geqslant \mathcal{C}(\Phi(\rho))$. 

(C2b). Monotonicity for average coherence under selective outcomes of ICPTP: $\mathcal{C}(\rho) \geqslant$ $\sum_{n} p_{n} \mathcal{C}\left(\rho_{n}\right)$, where $\rho_{n}=K_{n} \rho K_{n}^{\dagger} / p_{n}$ and $p_{n}=\operatorname{Tr}\left[K_{n} \rho K_{n}^{\dagger}\right]$ for all $K_{n}$ with $\sum_{n} K_{n}^{\dagger} K_{n}=\mathbf{1}$ and $K_{n} \mathcal{I} K_{n}^{\dagger} \subseteq \mathcal{I}$. 

(C3). Convexity, i.e. $\lambda \mathcal{C}(\rho)+(1-\lambda) \mathcal{C}(\sigma) \geqslant \mathcal{C}(\lambda \rho+(1-\lambda) \sigma)$, for any density matrix $\rho$ and $\sigma$ with $0 \leqslant \lambda \leqslant 1$.

A general distance-based coherence quantifier can be found \cite{Coh9}, which satisfies all the conditions mentioned above, regarding the minimal distance between {the} target state and a given incoherent state. Using quantum relative entropy
\be
S(\rho \| \sigma)=\operatorname{Tr}\left[\rho \log \rho\right]-\operatorname{Tr}\left[\rho \log \sigma\right], \label{eq14}
\ee
one coherence monotone is $\min _{\sigma \in \mathcal{I}} S(\rho \| \sigma)$ which can be recasted into \cite{Coh1}
\be
\mathcal{C}_{\text {R.E. }}({\rho})=S\left({\rho}_{\text {diag }}\right)-S({\rho}), \label{eq13}
\ee
where $S({\rho})=-\operatorname{Tr}({\rho} \log {\rho})$ is the von Neumann entropy for the state ${\rho}=\sum_{i j} \rho_{i j}\left|i\right\rangle\left\langle j\right|$, and ${\rho}_{\text {diag }}=\sum_{i} \rho_{i i}\left| i\right\rangle\left\langle i\right|$ is derived from a dephasing operation on a density matrix. 

In the next section, we will investigate {the} above coherence monotone for {a} two-UDW detector system. We should remark that we do not use another well-known $\l_1$-norm monotone as authors of \cite{Coh-U1,Coh-U2} did, because monotone (\ref{eq13}) is a more general measure even for {the} infinite-dimensional system to avoid potential divergent \cite{Coh-U3}. Although, in {the} present case, these two monotones are equivalent, we remain using (\ref{eq13}) for possible extension in future work.

\subsection{Coherence revival}

In our context, by diagonalizing the density matrix (\ref{eq11}), we obtain the related eigenvalues
\bea
\lambda_1=\frac{(1-\gamma)^2(3+\Delta_0)}{4\left(3+\gamma^2\right)}~&,&~
\lambda_2=\frac{(1+\gamma)^2(3+\Delta_0)}{4\left(3+\gamma^2\right)},\no\\
\lambda_3=\frac{(1-\gamma^2)(3+\Delta_0)}{3+\gamma^2}~&,&~
\lambda_4=\frac{1-\Delta_0}{4},   \label{eq15}
\eea
 and eigenvectors 
 \bea
 |\lambda_1\ra=|00\ra~~~&,&~~~ |\lambda_2\ra=|11\ra,\no\\
 |\lambda_3\ra=\frac{|10\ra-|01\ra}{\sqrt{2}}~~~&,&~~~ |\lambda_4\ra=\frac{|01\ra-|10\ra}{\sqrt{2}} . \label{eq16}
 \eea

We straightforwardly calculate the relative entropy as
\be
\mathcal{C}_{R.E.}(\omega,T_U,\Delta_0)
=-2C\log C+\lambda_3\log\lambda_3+\lambda_4\log\lambda_4,  \label{eq17}
\ee
which depends on the energy spacing of atom $\omega$, the initial state choice of detectors encoded in $\Delta_0$, as well as Unruh temperature $T_U$ proportional to the detectors' acceleration.

\begin{figure}[t]
\begin{center}
\includegraphics[width=.47\textwidth]{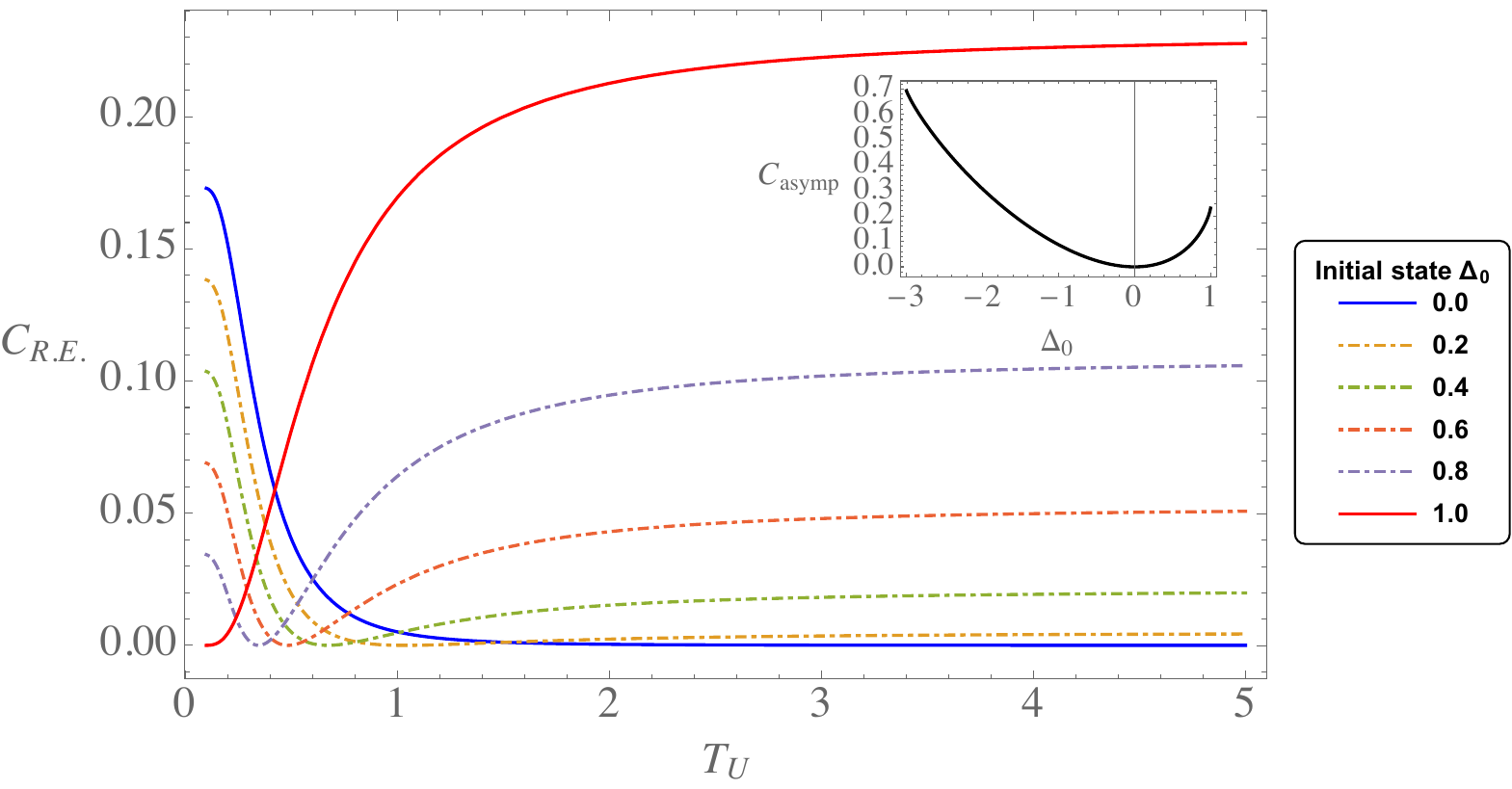}
\caption{When two-detector system evolving to an equilibrium, its quantum coherence monotone $\mathcal{C}_{R.E.}$ is plotted as a function of initial state choice $\Delta_0$ and Unruh temperature $T_U$, for fixed $\omega=1$. {A coherence revival phenomenon occurs for particular chosen initial states} with $\Delta_0\in(0,1]$. That is, as Unruh temperature $T_U$ {grows}, $\mathcal{C}_{R.E.}$ degrades at first to zero and then revives to a nonvanishing asymptotic value. In the inset, the dependence of asymptotic quantum coherence $\mathcal{C}_{\text{asymp}}$ on the choice of initial state is demonstrated.}
\label{fig1}
\end{center}
\end{figure}

Since $\omega$ and $T_U$ are combined in $\gamma$ (\ref{eq9}), for simplicity, we specify $\omega=1$ and depict the related quantum coherence in Fig.\ref{fig1}, as a function of initial state choice $\Delta_0$ and Unruh temperature $T_U$. Since {the} Unruh effect, in general, is recognized as an environment decoherence \cite{TAKAGI1}, we may naively expect that for arbitrary fixed initial state, quantum coherence $\mathcal{C}_{R.E.}$ would be a monotonously decreasing function of Unruh temperature $T_U$. However, we observe directly from Fig.\ref{fig1} that the quantum coherence is a non-monotonous function of $T_U$ {corresponding to a class of initial states}. In particular, there is a revival of coherence, i.e., as Unruh temperature $T_U$ {grows}, the quantum coherence degrades to zero at first and then increases to a nonvanishing asymptotic value $\mathcal{C}_{\text{asymp}}$, obtained from (\ref{eq17}) by taking the infinite acceleration limit $\gamma\rar0$. 

\begin{figure*}[t]
\begin{center}
\begin{minipage}{\textwidth}
\subfloat[$\mathcal{C}_{\text {R.E. }}(\omega=1)$]{\includegraphics[width=.31\textwidth]{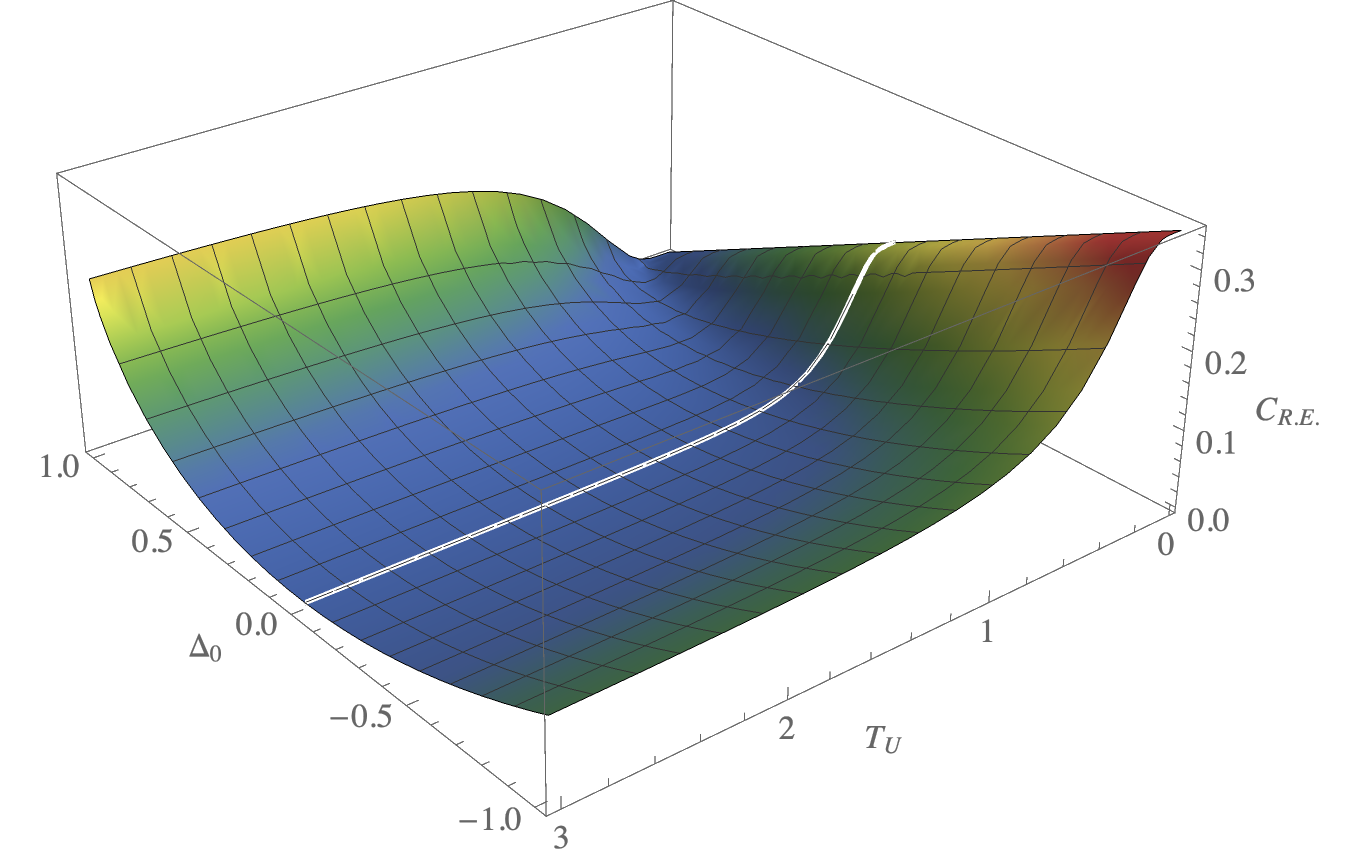}}~
\subfloat[$\mathcal{C}_{\text {R.E. }}(\omega=3)$]{\includegraphics[width=.28\textwidth]{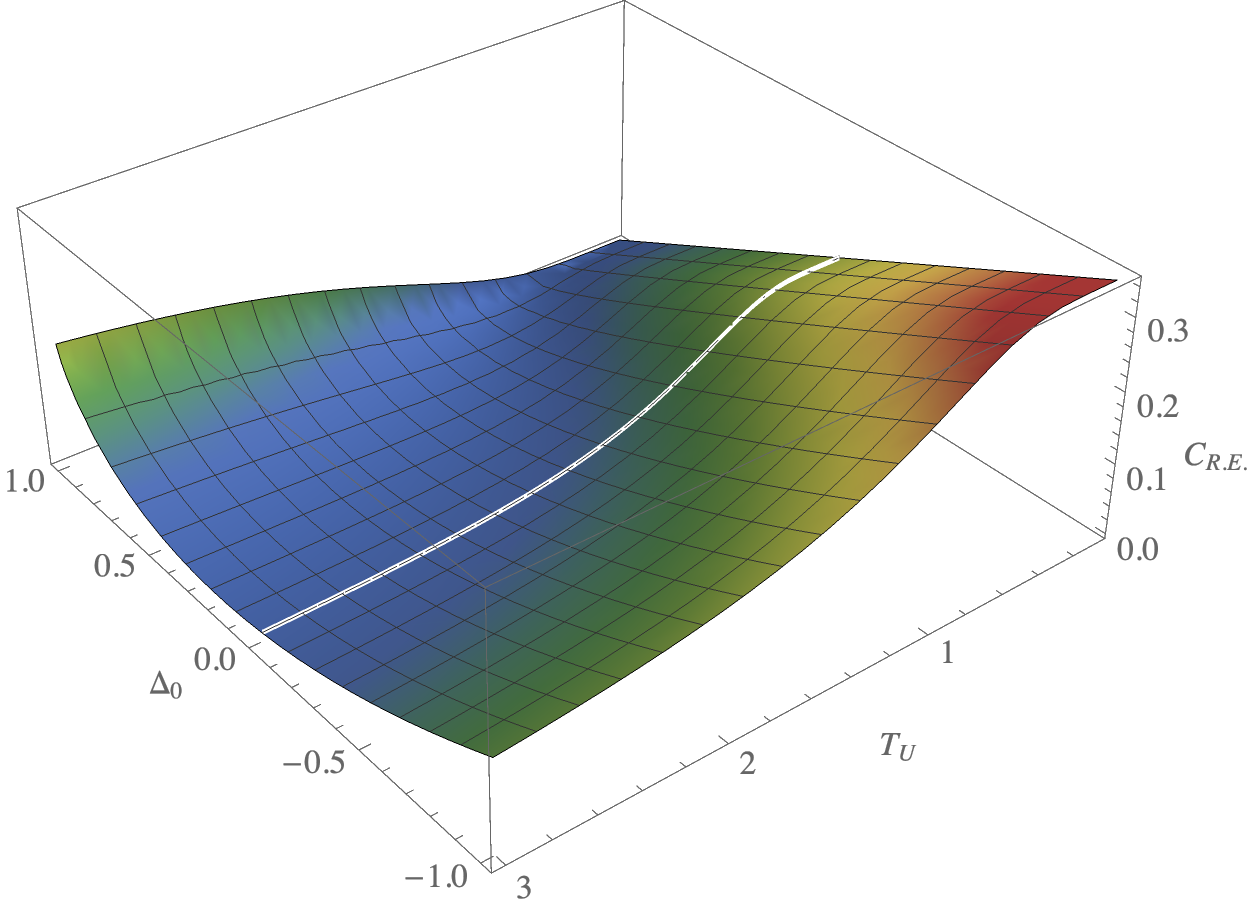}}~
\subfloat[$\mathcal{C}_{\text {R.E. }}(\omega=5)$]{\includegraphics[width=.30\textwidth]{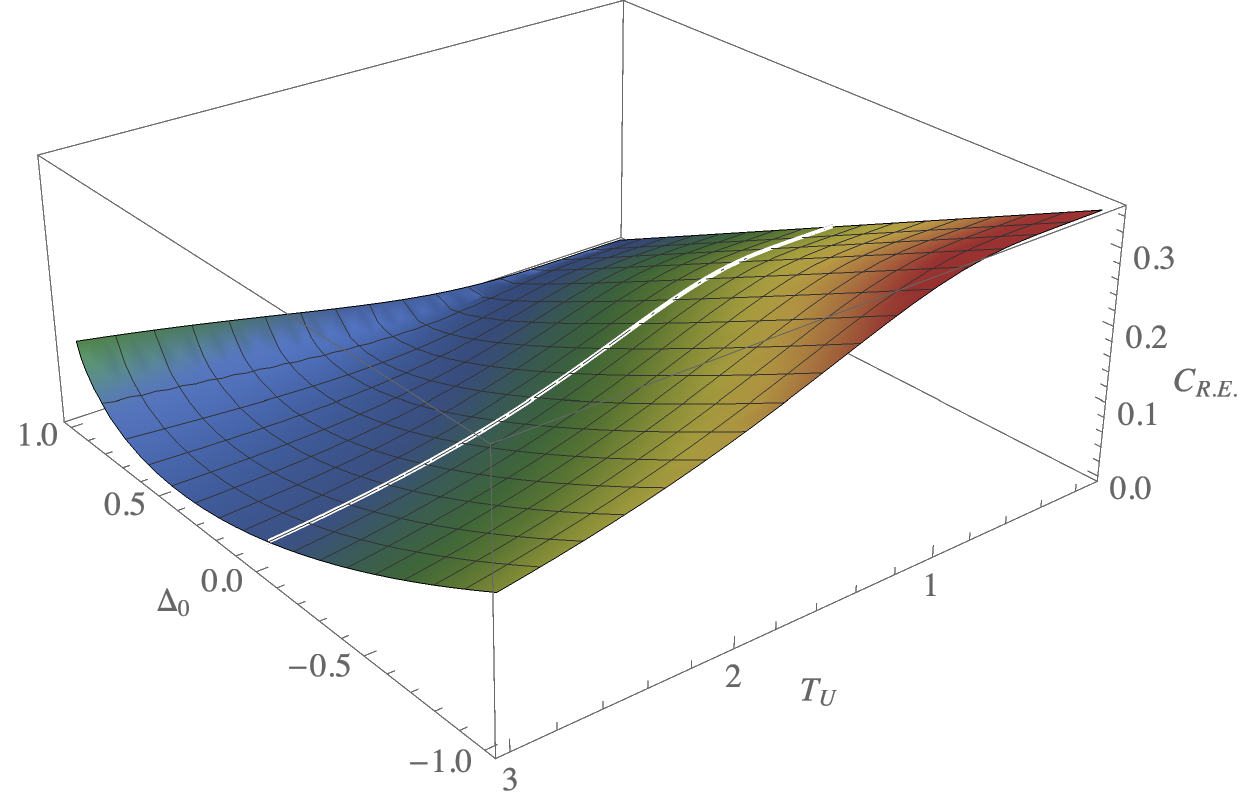}}\\
\subfloat[$\mathcal{E}_{\text {neg}}(\omega=1)$]{\includegraphics[width=.30\textwidth]{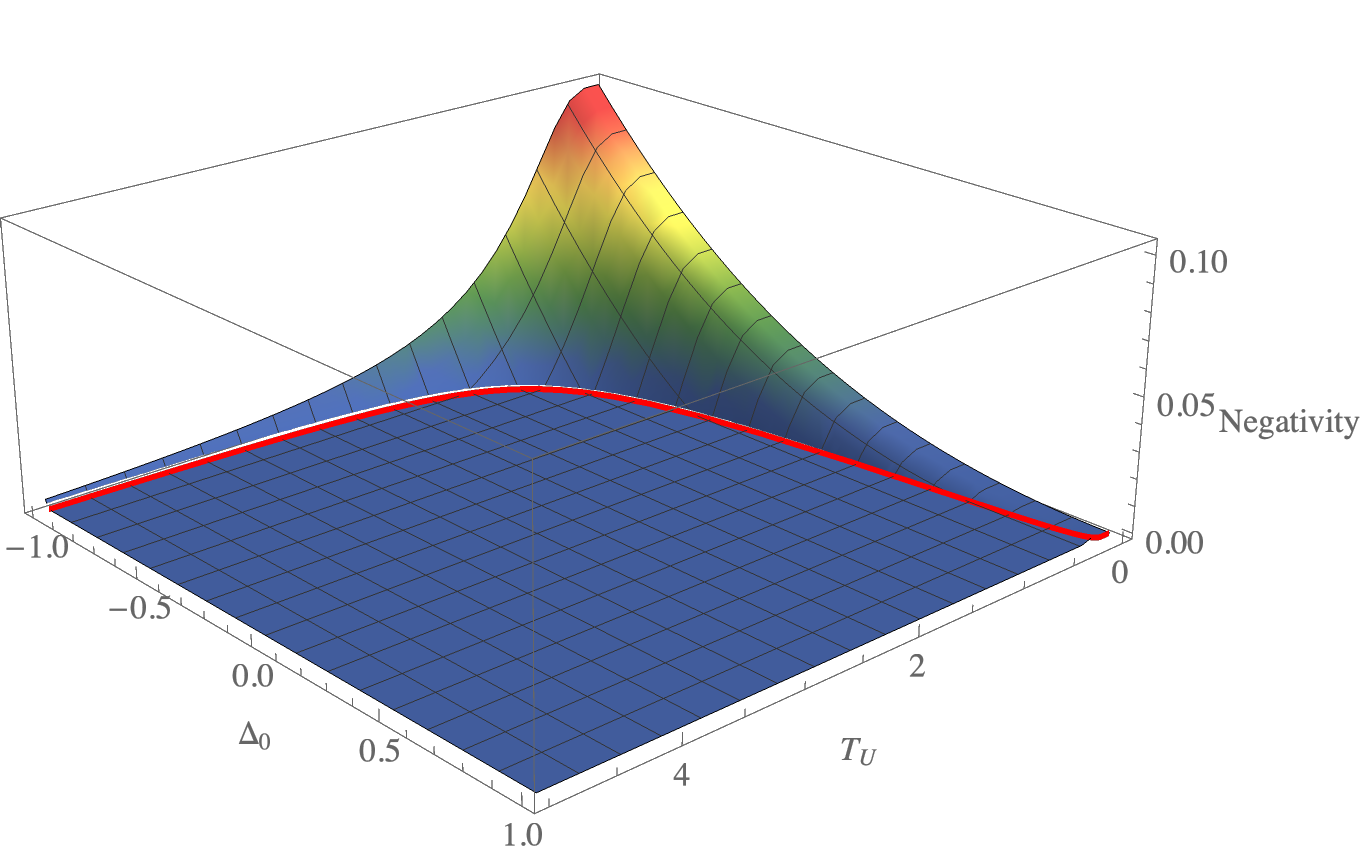}}
\subfloat[$\mathcal{E}_{\text {neg}}(\omega=3)$]{\includegraphics[width=.33\textwidth]{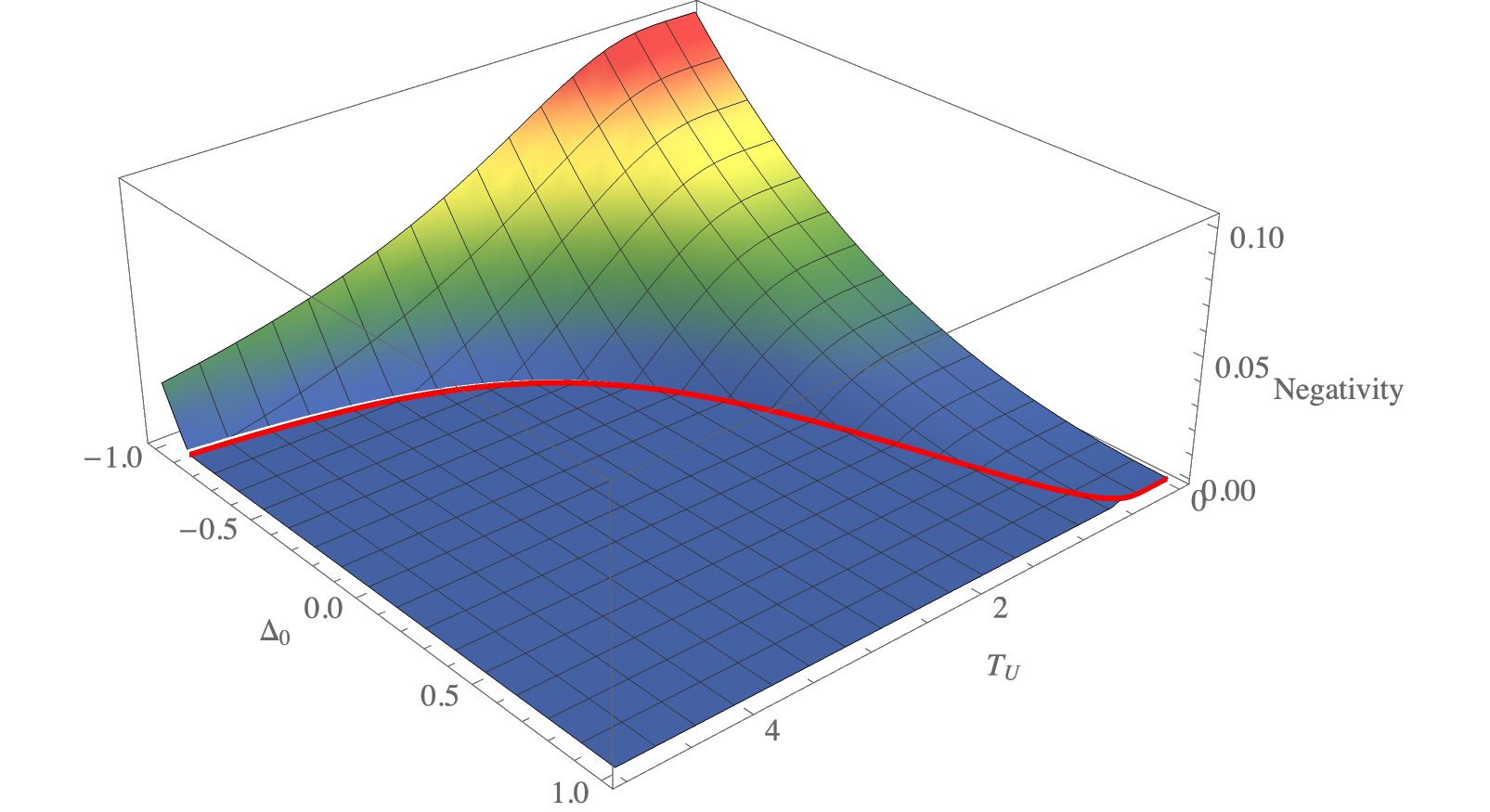}}
\subfloat[$\mathcal{E}_{\text {neg}}(\omega=5)$]{\includegraphics[width=.30\textwidth]{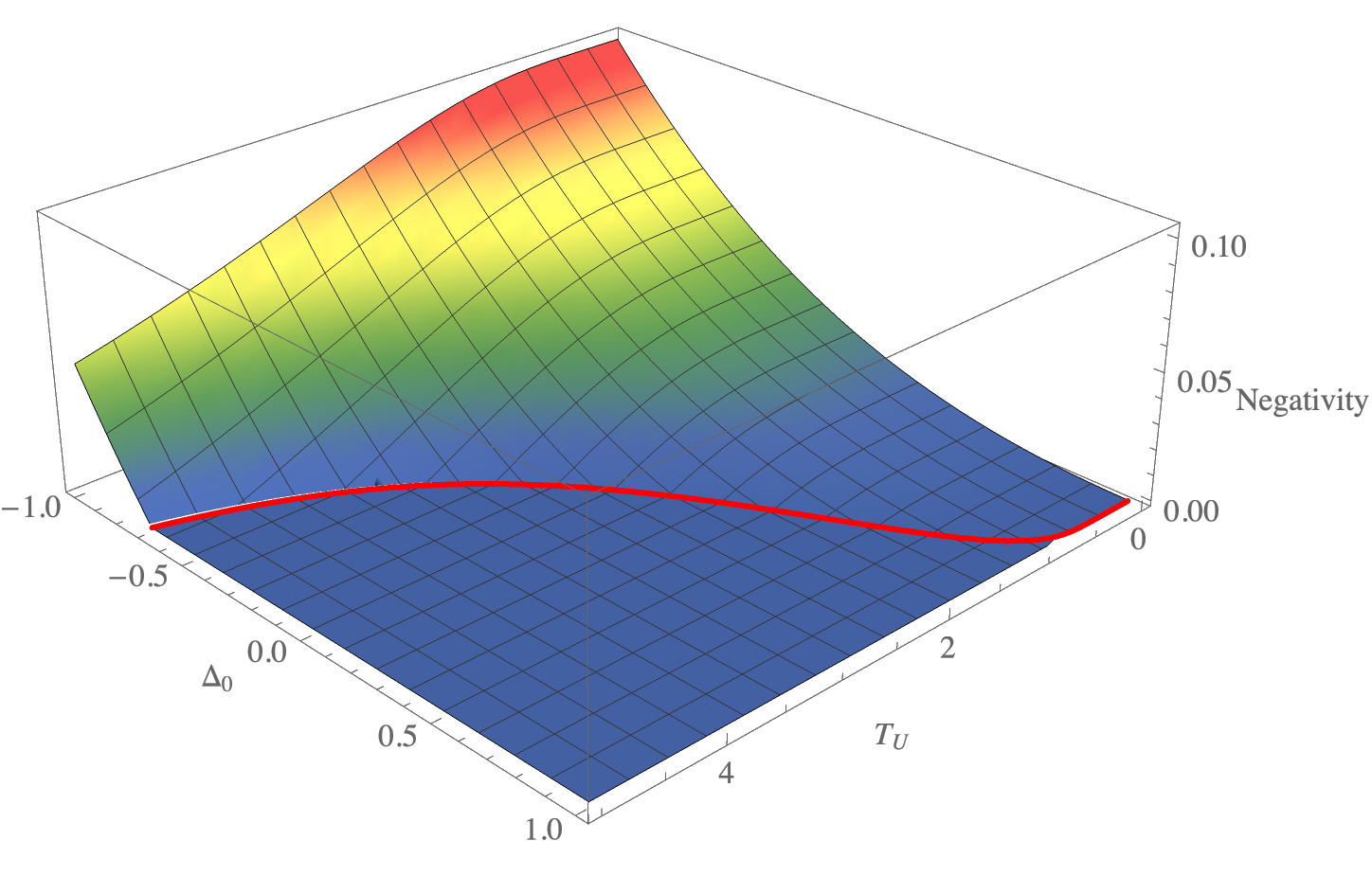}}
\end{minipage}
\caption{(Upper) The quantum coherence of the final equilibrium state of two-detectors system as a function of initial state preparation encoded by $\Delta_0$ and Unruh temperature $T_U$. We demonstrate the coherence monotone $\mathcal{C}_{R.E.}$ for fixed energy spacing of detector with $\omega=1,3,5$. In each case, coherence revival can exist for the initial states with $\Delta_0\in(0,1]$. (Lower) Entanglement measured by negativity is plotted as a function of initial state preparation encoded by $\Delta_0$ and Unruh temperature $T_U$. For any fixed initial states, no revival of entanglement with respect to Unruh temperature can occur.}
\label{fig2}
\end{center}
\end{figure*}

In the upper line of Fig.\ref{fig2}, we plot the quantum coherence as a function of initial state choice $\Delta_0$ and Unruh temperature $T_U$, for different {values of the} detector's energy spacing. By {numerical} analysis, we find that above \emph{coherence revival} occurs when the initial state satisfying $\Delta_0\in(0,1]$, bounded by the white solid curves in Fig.\ref{fig2}(a)-\ref{fig2}(c). We also note that the revival of coherence is suppressed for {the} larger energy spacing of a UDW detector. 


A striking difference between quantum coherence and entanglement generated under {the} Unruh effect \cite{Entg6} needs to be highlighted. It has been shown that the {survival} of entanglement for the equilibrium state of two accelerating detectors {results from} competition between the Unruh decoherence and entanglement enhancement from Markovian dynamics. In other words, during the process of entanglement generation, {the} Unruh effect solely plays a role {in} decoherence; therefore, as {the} Unruh temperature {grows}, the survived entanglement degrades. To see this, we directly calculate the negativity for the final equilibrium state of two accelerating UDW detectors, a measure of distillable entanglement. It is defined by $\mathcal{E}_{\text{neg}}(\rho)=\frac{1}{2} \sum_{i}\left(\left|\lambda_{i}\right|-\lambda_{i}\right)=-\sum_{\lambda_{i}<0} \lambda_{i}$, where $\lambda_{i}$ are the negative eigenvalues of partial transposed density matrix. The value of negativity ranges from 0 for a separable state, to $0.5$ for a maximally entangled state. Form (\ref{eq11}), we can straightforwardly obtain \cite{F1}
\be
\mathcal{E}_{\text{neg}}=\max\left\{\frac{ \sqrt{\left(\Delta_0-\gamma^{2}\right)^{2}+\gamma^{2}\left(3+\Delta_0\right)^2}}{2\left(3+\gamma^{2}\right)}-\frac{\lambda_1+\lambda_2}{2},0\right\}, \label{eq18}
\ee
which reaches maximum $0.5$ at $\Delta_0=-3$. In the lower line of Fig.\ref{fig2}, we plot the negativity for the final equilibrium state of {the} two-detector system for $\omega=1,3,5$. For varying initial state preparation, the negativity (\ref{eq18}) is vanishing at $\Delta_0=\frac{5 \gamma^{2}-3}{3-\gamma^{2}}$, presented by the red solid curves in Fig.\ref{fig2}(d)-\ref{fig2}(f). It is obvious that entanglement degrades monotonously with growing Unruh temperature, i.e., no revival of entanglement can happen for any initial state preparation. This clearly demonstrates the distinction between entanglement and coherence \cite{Coh7}.

\subsection{A concrete example}
We illustrate our result in a concrete example. Choosing the initial state of two-detector in a product form
\be
\rho_{\text{initial}}=\rho_A(0)\otimes\rho_B(0) .     \label{eq19}
\ee 
The state of each detector can be written in Bloch form
\be
\rho_{A}(0)=\frac{1}{2}\left(\mathbf{1}+\bm{n}\cdot\bm{\sigma}\right),~~~\rho_{B}(0)=\frac{1}{2}\left(\mathbf{1}+\bm{m}\cdot\bm{\sigma}\right) ,  \label{eq20}
\ee
where $\bm{n}$ and $\bm{m}$ are two unit Bloch vectors. Without loss of generality, taking $\bm{n}=(0,0,1)$ and $\bm{m}=(0,\sin\theta,\cos\theta)$, we have $\Delta_0=\bm{n}\cdot\bm{m}=\cos\theta$ where $\theta\in[0,\pi]$ is angle between two vectors giving $\Delta_0\in[-1,1]$. 

The quantum coherence of the initial state (\ref{eq19}) is 
\be
\mathcal{C}_{R.E.}(0)=H_{\text{binary}}\left(\frac{\cos\theta+1}{2}\right) , \label{eq21}
\ee 
where $H_{\text{binary}}(x)\equiv-x\log x-(1-x)\log(1-x)$ is binary entropy of variable $x$.

After the Markovian evolution, the final equilibrium state of two detectors {possesses} quantum coherence (\ref{eq17}). Defining the change between the final state coherence and those of {the} detector initially as
\be
\delta\mathcal{C}_{R.E.}=\mathcal{C}_{R.E.}(\omega,T_U,\theta)-\mathcal{C}_{R.E.}(0).  \label{eq22}
\ee
Once $\delta\mathcal{C}_{R.E.}>0$ (i.e., more coherence for detectors' final state than it was had initially), the quantum coherence has been generated through the interaction between the detectors' system and the background field. By designed operation protocol, one may expect {such coherence increment to} be properly extracted. 

\begin{figure}[t]
\begin{center}
\includegraphics[width=.39\textwidth]{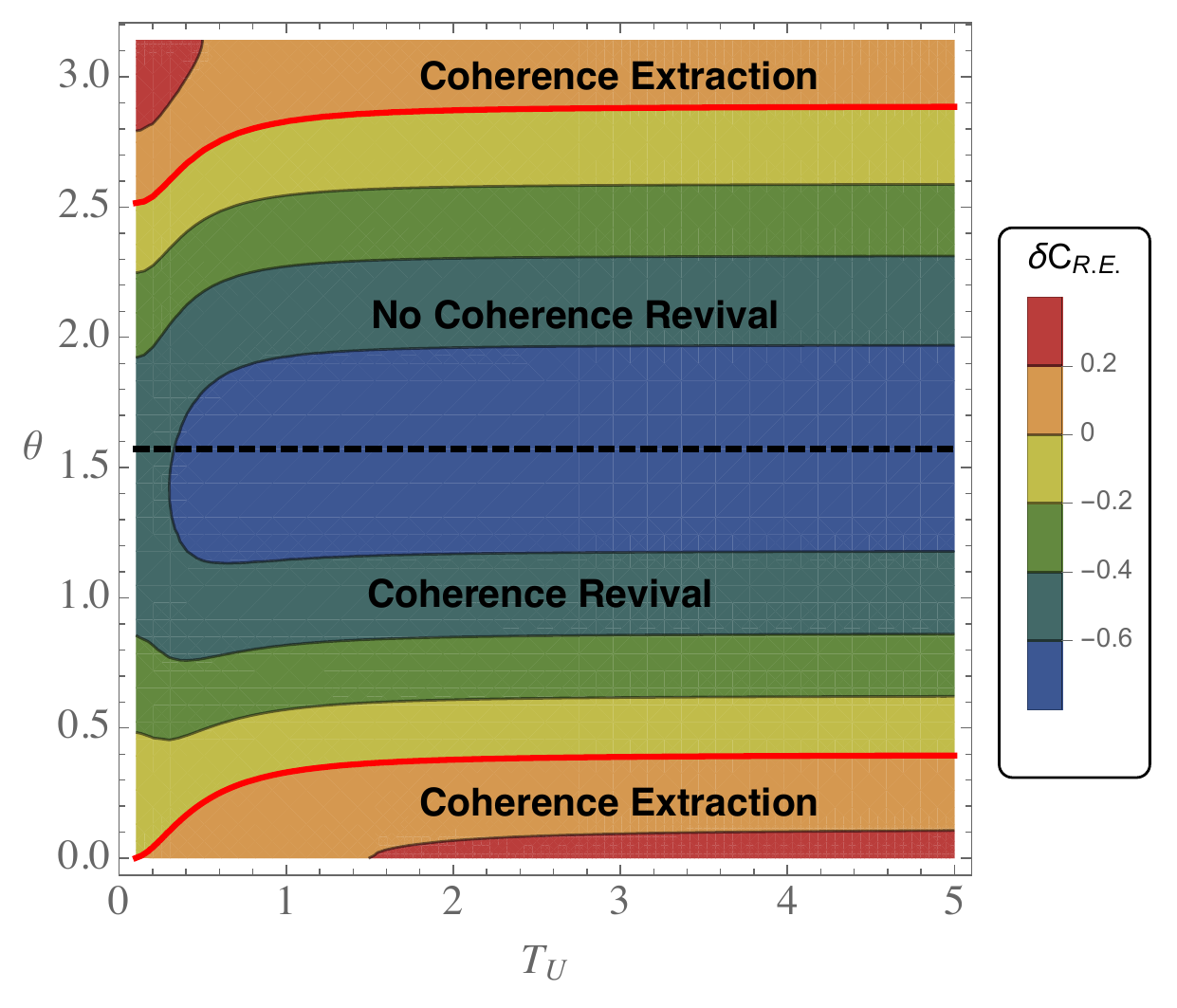}
\caption{The quantum coherence extraction for two UDW detectors (fixed $\omega=1$). For initial states preparation with $\theta\in[0,\frac{\pi}{2}]$, coherence revival occurs for sufficiently large Unruh temperature. Further, for initial states with $\theta$ chosen inside the regions bounded by the red solid lines, detectors can possess more quantum coherence at equilibrium than {their} initial state, which means a coherence extraction can happen.}
\label{fig3}
\end{center}
\end{figure}

Numerically, we depict $\delta\mathcal{C}_{R.E.}$ in Fig.\ref{fig3}. Starting from an incoherent initial state with $\theta=0,\pi$, one has $\mathcal{C}_{R.E.}(0)=0$ but a nonvanishing $\delta\mathcal{C}_{R.E.}>0$ for two-detector at final equilibrium, which indicates a generation of quantum coherence through its evolution. For general initial states, we know that a coherence revival can occur with choice $\theta\in[0,\pi/2)$ (i.e., $\Delta_{0} \in(0,1]$) for sufficiently large Unruh temperature. However, by numerically evaluating, we recognize only regions bounded by the red solid lines in Fig.\ref{fig3} can have $\delta\mathcal{C}_{R.E.}>0$, referring to a coherence extraction for sufficiently large $T_U$, i.e., the coherence of two-detector's final state is enhanced compared to its initial state. On the other hand, the region in Fig.\ref{fig3} without coherence extraction indicates that the generated coherence via Markovian evolution of the detectors cannot counteract the effect of Unruh decoherence.

\section{Metrological advantage}
\label{4}

\begin{figure*}[t]
\begin{center}
\begin{minipage}{\textwidth}
\subfloat[$\mathcal{F}_{T}(\omega=1)$]{\includegraphics[width=.31\textwidth]{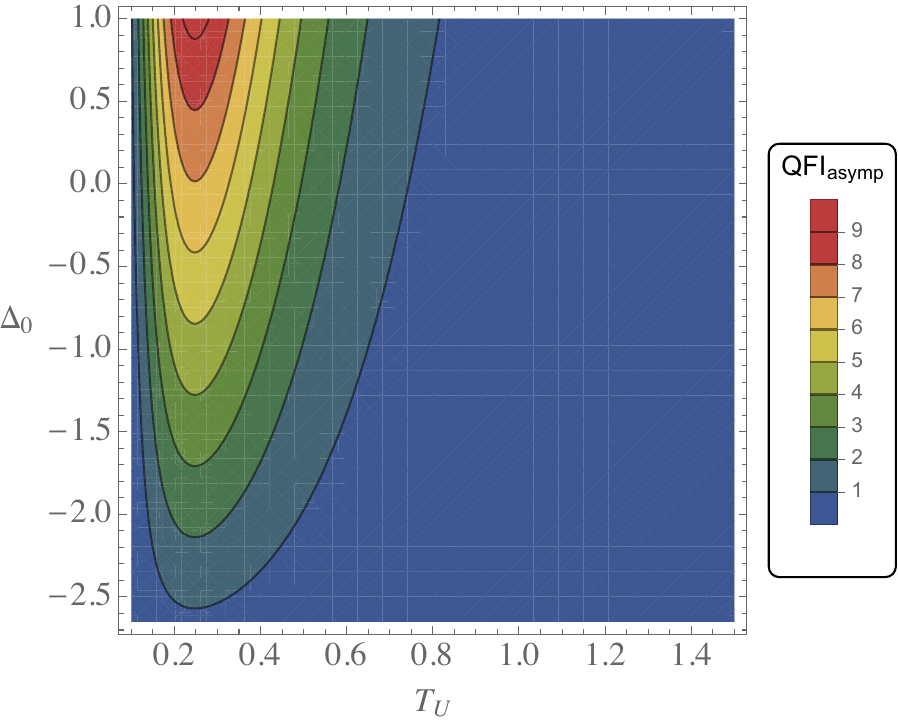}}~
\subfloat[$\mathcal{F}_{T}(\omega=3)$]{\includegraphics[width=.31\textwidth]{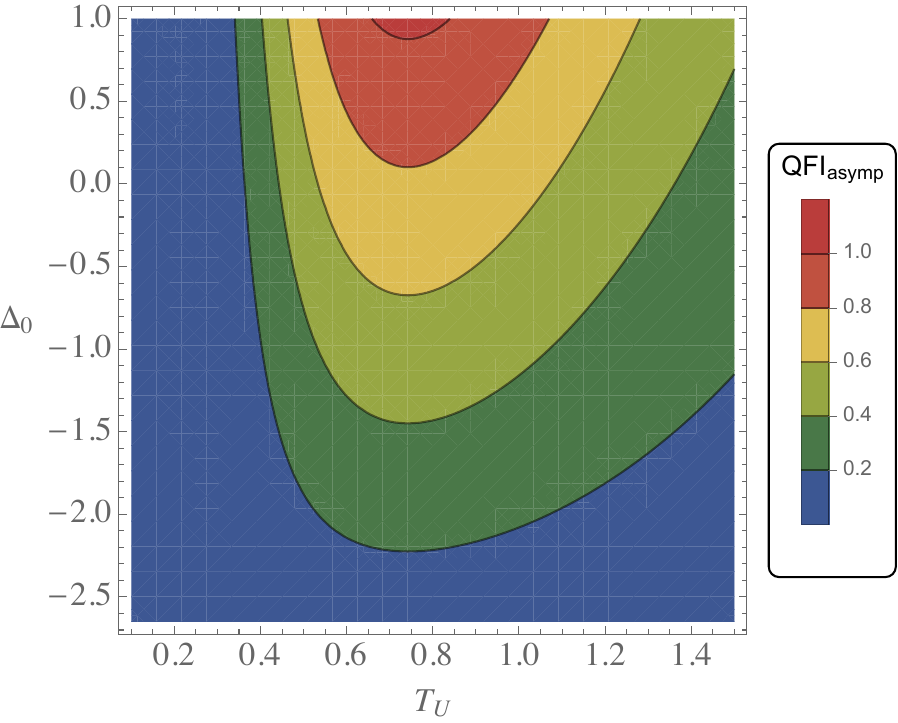}}~
\subfloat[$\mathcal{F}_{T}(\omega=5)$]{\includegraphics[width=.31\textwidth]{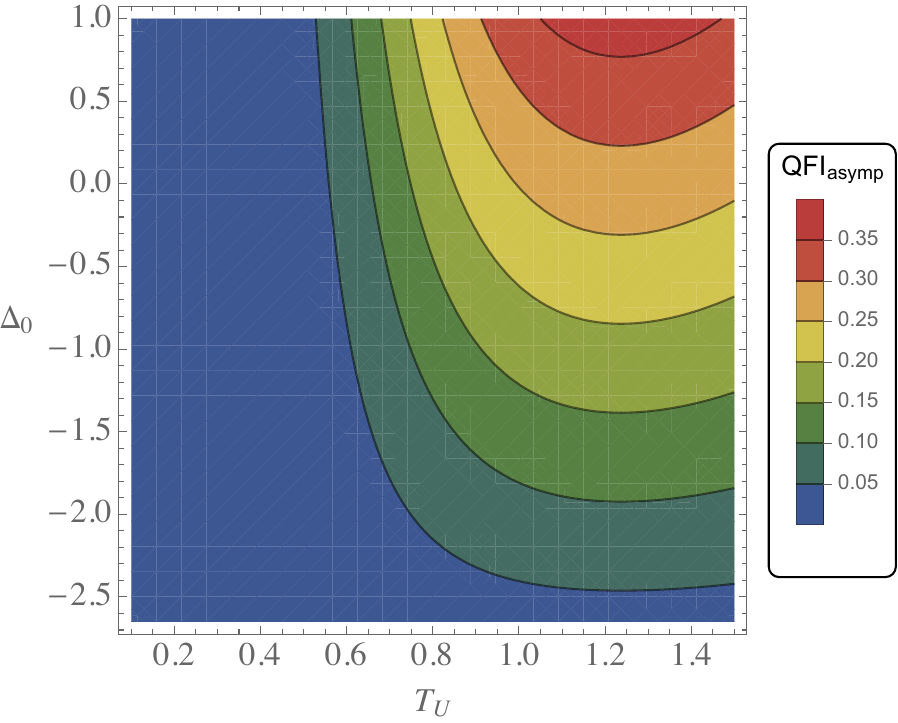}}
\end{minipage}
\caption{The quantum Fisher information (QFI) of the final equilibrium state of two-detectors system as a function of initial state preparation encoded by $\Delta_0$ and Unruh temperature $T_U$. We choose Unruh temperature as a parameter to be estimated by quantum measurement. The related QFI $\mathcal{F}_{T}$ is demonstrated for fixed energy spacing of {the} detector as $\omega=1,3,5$. In each case, {a} local peak of QFI can exist, with its maximal value significantly enhanced in a range of $\Delta_0$ where coherence revival occurs. As {the} revival of quantum coherence is suppressed for larger $\omega$, the maximal value of QFI also degrades significantly.}
\label{fig4}
\end{center}
\end{figure*}

In general, quantum coherence is recognized as a physical resource that can be utilized to improve the performance of various quantum information tasks. One particular example is the close relation between non-classical coherence and quantum metrological tasks, where the quantumness encoded in coherence can be used to enhance the accuracy of estimation of extremely sensitive {parameters}, e.g., Unruh temperature, spacetime curvature in relativistic context \cite{QFIadd1,QFIadd2,Wang,Tian,F4,QFIadd3}. {This section aims} to examine the metrological advantage of coherence revival found before. In particular, we will evaluate the QFI for the two-UDW detector, with Unruh temperature as a parameter chosen to be optimally estimated. We will show that the revival coherence corresponds to the enhanced QFI {and,} therefore, may provide a significant quantum advantage in metrological {tasks}.
   
In any quantum metrological task, QFI gives a lower bound to the mean-square error in the parameter estimation via the Cram\'er-Rao inequality $\operatorname{Var}(X) \geqslant\left[N \mathcal{F}_{X}\right]^{-1}$, where $N$ is the number of repeated measurements. In terms of the symmetric logarithmic derivative (SLD) operator $L_X$, which satisfies $\p_X\rho=\frac{1}{2}\{\rho,L_X\}$, the QFI is defined as $\mathcal{F}_X=\mbox{Tr}[\rho(X)L^2_X]$. For a diagonalized density matrix like $\rho=\sum_{i=\pm}\lambda_i|\psi_i\ra\la\psi_i|$, the related QFI can be further written as \cite{QFI1,QFI2}
\be
\mathcal{F}_X=\sum_{i=\pm}\frac{(\p_X \lambda_i)^2}{\lambda_i}+\sum_{i\neq j=\pm}\frac{2(\lambda_i-\lambda_j)^2}{\lambda_i+\lambda_j}|\la\psi_i|\p_X\psi_j\ra|^2 ,           \label{QFI}
\ee
where the summations run over all eigenvalues satisfying $\lambda_i\neq0$ and $\lambda_i+\lambda_j\neq0$.

We choose to estimate Unruh temperature $T_U$. The related QFI can be straightforwardly calculated from (\ref{QFI}) by substituting the eigenvalues (\ref{eq15}) of {the} diagonalized density matrix of {the} detectors system. According to different preparations of initial state encoded in $\Delta_0\in[-3,1]$, we come to two classes of QFI:

(i) $\Delta_0=-3$, only one nonvanishing eigenvalue $\lambda_4=1$, which gives $\mathcal{F}_{T}=0$;

(ii) $\Delta_0\in(-3,1]$, all four eigenvalues (\ref{eq15}) are nonvanishing. Since $\p_T\lambda_4=0$, we have the QFI 
\bea
\mathcal{F}_{T}&=&\sum_{i=1,2,3}\frac{(\p_{T}\lambda_i)^2}{\lambda_i}\no\\
&=&\frac{ \gamma^{6}-9 \gamma^{4}-\gamma^{2}+9}{\gamma^{6}+9 \gamma^{4}+27 \gamma^{2}+27}\cdot\frac{\omega^2(3+\Delta_0)}{2T_U^4}.
\eea

With specific energy spacing of detector $\omega=1,3,5$, we depict the QFI in Fig.\ref{fig4} as a function of Unruh temperature and initial state choice $\Delta_0$. The first thing {that} steps out is that similar to quantum coherence rather than entanglement, with fixed initial state choice, the QFI of {the} two-UDW detector is not a monotonous function of Unruh temperature. We note that a local peak exists as Unruh temperature {grows}, which means the largest precision of estimation on $T_U$ can be achieved at {relatively} low acceleration. Such non-monotonicity of the QFI with respect to Unruh temperature was also found in {the} single-detector case recently \cite{nomono1}.

Comparing Fig.\ref{fig4} and the upper line of Fig.\ref{fig2}, we find further similarity between quantum coherence and the QFI. It is {obvious} that the maximal value of the QFI peak can be achieved for initial state preparation with $\Delta_0=1$, i.e., as approaching $\Delta_0=1$, the increasing of QFI corresponds to enhanced quantum coherence revival. For larger $\omega$, the coherence revival is heavily suppressed, and correspondingly, we find that the maximal value of {the} QFI peak also decreases conspicuously.   

In general, the quantumness of {a} physical system can be utilized as a physical resource in various quantum processes. Although both quantum coherence and entanglement can serve as a resource, due to the complementary non-monotonous behaviors shared by coherence and QFI under growing $T_U$, we suggest that it is the coherence rather than the entanglement dominates the QFI's local peak and the maximal value. In other words, with coherence revival, one can significantly improve the precision in estimating Unruh temperature, thus endowing quantum coherence a metrological advantage.

\section{Summary and discussions}
\label{5}
In this paper, we explore the quantum coherence generation between two accelerating Unruh-DeWitt detectors. We find coherence can be generated through the Markovian detector evolution, just like entanglement dynamics in {the} same scenario. While {the monotonous generation} of quantum entanglement is well-known, we find that for certain {choices} of detectors' initial state, under growing Unruh temperature, coherence measure may have non-monotonous behavior. We verify such coherence revival by inspecting its advantage in enhancing the QFI of Unruh temperature estimation. As the maximal QFI {bounds} the accuracy of quantum measurement, we conclude that the extracted coherence may be utilized as a physical resource in quantum metrology. 

It seems curious at first glance that coherence monotone (\ref{eq13}) and quantum entanglement have distinctive {behaviors} under Unruh decoherence, as many works have indicated \cite{Coh9} that both of them are physical resources that {are} interchangeable {with} each other. However, we note that, unlike entanglement, coherence is a basis-dependent concept. If we adopt the viewpoint that a preferred basis of coherence only emerges by the environment (e.g., the energy conservation makes the energy eigenbases naturally selected in the paper as preferred bases), with varying Unruh temperature, we would better cautiously refer {to} the related background states as {the} different environment. In this context, the coherence revival could be a demonstration of so-called ''einselection'' \cite{Disc1} in {a} relativistic framework. 

On the other hand, the intimate relation between entanglement dynamics and coherence generation of multi-UDW detectors may be reexamined from a thermodynamic perspective. In particular, for an open quantum system, its evolution irreversibility can be characterized by the entropy production arising from non-equilibrium quantum processes ascribed to system correlations. It has been recently shown \cite{Disc2} that the entropy production of an open quantum system can further be interpreted as an interplay between population dynamics and coherence dynamics. Extending this analysis to the context of Unruh decoherence may shed new light on our understanding {of} the quantum nature of {the} Unruh effect. We will exploit these interesting topics in {the} future.

\section*{Acknowledgement}
This work is supported by the National Natural Science Foundation of China (Nos.12075178, 12475061), Shaanxi Fundamental Science Research Project for Mathematics and Physics (No. 23JSY006), and the Innovation Program for Quantum Science and Technology (2021ZD0302400). 



\begin{thebibliography}{999}



\bibitem{Entg1}
R. Horodecki, P. Horodecki, M. Horodecki, and K. Horodecki, Quantum entanglement, Rev. Mod. Phys. \textbf{81}, 865 (2009).

\bibitem{Entg2}
D. Harlow, Jerusalem lectures on black holes and quantum information, Rev. Mod. Phys. \textbf{88}, 015002 (2016).

\bibitem{Entg3}
N. Brunner, D. Cavalcanti, S. Pironio, V. Scarani and S. Wehner, Bell nonlocality, Rev. Mod. Phys. \textbf{86}, 419 (2014).

\bibitem{Entg4}
H. Ollivier and W. H. Zurek, Quantum Discord: A Measure of the Quantumness of Correlations, Phys. Rev. Lett. \textbf{88}, 017901 (2001).

\bibitem{Coh1}
T. Baumgratz, M. Cramer, and M. B. Plenio, Quantifying Coherence, Phys. Rev. Lett. \textbf{113}, 140401 (2014).

\bibitem{Coh2}
A. Winter and D.Yang, Operational Resource Theory of Coherence, 
Phys. Rev. Lett. \textbf{116},120404 (2016).

\bibitem{Coh3}
T. R. Bromley, M. Cianciaruso, and G. Adesso, Frozen Quantum Coherence, 
Phys. Rev. Lett. \textbf{114}, 210401 (2015).

\bibitem{Coh4}
I. Marvian and R. W. Spekkens, How to quantify coherence: Distinguishing speakable and unspeakable notions, Phys. Rev. A \textbf{94}, 052324 (2016).

\bibitem{Coh5}
S. Lloyd, Quantum coherence in biological systems, J. Phys. Conf. Ser. \textbf{302}, 012037 (2011).

\bibitem{Coh6}
J. \AA berg, Catalytic Coherence, Phys. Rev. Lett. \textbf{113}, 150402 (2014).

\bibitem{Coh7}
A. Streltsov, U. Singh, H. S. Dhar, M. N. Bera, and Adesso, Measuring Quantum Coherence with Entanglement, Phys. Rev. Lett. \textbf{115}, 020403 (2015).

\bibitem{Coh8}
G. E. Chitambar and M.-H. Hsieh, Relating the Resource Theories of Entanglement and Quantum Coherence, Phys. Rev. Lett. \textbf{117}, 020402 (2016).

\bibitem{Coh9}
A. Streltsov, G. Adesso, and M. B. Plenio, Quantum coherence as a resource, 
Rev. Mod. Phys. \textbf{89}, 041003 (2017).

\bibitem{Coh10}
M. Hu, X. Hu, J. Wang, Y. Peng, Y. R. Zhang, and H. Fan, Quantum coherence and geometric quantum discord, 
Phys. Rep. \textbf{762}, 1(2018).



\bibitem{Entg6}
F. Benatti and R. Floreanini, Entanglement generation in uniformly accelerating atoms: reexamination of the Unruh effect, 
Phys. Rev. A \textbf{70}, 012112 (2004).

\bibitem{Entg7}
A. Pozas-Kerstjens and E. Mart\'in-Mart\'inez, Harvesting correlations from the quantum vacuum, Phys. Rev. D \textbf{92}, 064042 (2015).




\bibitem{F1}
J. Feng, Y. Z. Zhang, M. D. Gould, and H. Fan, Uncertainty relation in Schwarzschild spacetime, 
Phys. Lett. B \textbf{743}, 198 (2015).

\bibitem{F2}
J. Feng, Y. Z. Zhang, M. D. Gould, and H, Fan, Fine-grained uncertainty relations under relativistic motion, 
Europhys. Lett. \textbf{122}, 60001 (2018).

\bibitem{Entg7+}
J. Hu and H. Yu, Entanglement dynamics for uniformly accelerated two-level atoms, 
Phys. Rev. A \textbf{91}, 012327 (2015).

\bibitem{F3}
J. Feng, X. Huang, Y. Z. Zhang, and H. Fan, Bell inequalities violation within non-Bunch-Davies states, 
Phys. Lett. B \textbf{786}, 403 (2018).

\bibitem{U1}
R. M. Wald, \emph{Quantum Field Theory in Curved Spacetime and Black Hole Thermodynamics} (University of Chicago Press, 1994).


\bibitem{Coh-U1}
N. K. Kollas, D. Moustos and K. Blekos, Field assisted extraction and swelling of quantum coherence for moving Unruh-DeWitt detectors, Phys. Rev. D \textbf{102}, 065020 (2020).

\bibitem{Coh-U2}
N. K. Kollas and D. Moustos, Generation and catalysis of coherence with scalar fields, Phys. Rev. D \textbf{105}, 025006 (2022).

\bibitem{Op1}
H.-P. Breuer and F. Petruccione, \emph{The Theory of Open Quantum Systems} (Oxford University Press, Oxford, 2002).

\bibitem{Entg8}
F. Benatti, R. Floreanini and U. Marzolino, Entangling two unequal atoms through a common bath, 
Phys. Rev. A \textbf{81}, 012105 (2010).

\bibitem{Entg9}
J. Hu and H. Yu, Entanglement dynamics for uniformly accelerated two-level atoms, 
Phys. Rev. A \textbf{91}, 012327 (2015).

\bibitem{metro1}
J. Sahota and N. Quesada, Quantum correlations in optical metrology: Heisenberg-limited phase estimation without mode entanglement, 
Phys. Rev. A \textbf{91}, 013808 (2015).

\bibitem{metro2}
H. Kwon, K. C. Tan, T. Volkoff, and H. Jeong, Nonclassicality as a Quantifiable Resource for Quantum Metrology, Phys. Rev. Lett. \textbf{122}, 040503 (2019).

\bibitem{Coh11}
L. Li , Q. -W. Wang, S. -Q. Shen, and M. Li, Quantum coherence measures based on Fisher information with applications, Phys. Rev. A.\textbf{103}, 012401 (2021).

\bibitem{Coh12}
M. Masini, T. Theurer, and M. B. Plenio, Coherence of operations and interferometry, Phys. Rev. A.\textbf{103}, 042426 (2021).

\bibitem{Coh13}
I. Marvian, Coherence distillation machines are impossible in quantum thermodynamics, Nat. Commun. \textbf{11}, 25 (2020).


\bibitem{nonmarkov}
D. Moustos and C. Anastopoulos, Non-Markovian time evolution of an accelerated qubit, Phys. Rev. D \textbf{95}, 025020 (2017).



\bibitem{Op2}
J. Hu and H. Yu, Quantum entanglement generation in de Sitter spacetime, Phys. Rev. D. \textbf{88}, 104003 (2013).

\bibitem{Op3}
V. Gorini, A. Kossakowski and E. C. G. Surdarshan, Completely positive dynamical semigroups of N-level systems, J. Math. Phys. \textbf{17}, 821 (1976).

\bibitem{Op4}
G. Lindblad, On the generators of quantum dynamical semigroups, Commun. Math. Phys. \textbf{48}, 119 (1976).


\bibitem{Plus1}
J. Hu and H. Yu, Entanglement generation outside a Schwarzschild black hole and the Hawking effect, J. High Energy Phys. \textbf{08}, 137 (2011).

\bibitem{Coh-U3}
Y.-R. Zhang, L.-H. Shao, Y. Li and H. Fan, Quantifying coherence in infinite-dimensional systems, Phys. Rev. A \textbf{93}, 012334 (2016).


\bibitem{TAKAGI1}
S. Takagi, Vacuum noise and stress induced by uniform accelerator: Hawking-Unruh effect in Rindler manifold of arbitrary dimensions, 
Prog. Theor. Phys. Suppl. \textbf{88}, 1 (1986).


\bibitem{QFIadd1}
M. Ahmadi, D. E. Bruschi, and I. Fuentes, Quantum metrology for relativistic quantum fields, Phys. Rev. D \textbf{89}, 065028 (2014).

\bibitem{QFIadd2}
M. Ahmadi, D. E. Bruschi, C. Sab\'in, G.Adesso, and I. Fuentes, Relativistic Quantum Metrology: Exploiting relativity to improve quantum measurement technologies, Sci. Rep. \textbf{4}, 4996 (2014).

\bibitem{Wang}
J. Wang, Z. Tian, J. Jing and H. Fan, Quantum metrology and estimation of Unruh effect, Sci. Rep. \textbf{4}, 7195 (2014).

\bibitem{Tian}
Z. Tian, J. Wang, H. Fan and J. Jing, Relativistic Quantum Metrology in Open System Dynamics, Sci. Rep. \textbf{5}, 7946 (2015).

\bibitem{F4}
X. Huang, J. Feng, Y. Z. Zhang, and H. Fan, Quantum estimation in an expanding spacetime, 
Ann. Phys. \textbf{397}, 336 (2018).


\bibitem{QFIadd3}
A. Lapponi, D. Moustos, D. E. Bruschi, and S. Mancini, Relativistic quantum communication between harmonic oscillator detectors, Phys. Rev. D \textbf{107}, 125010 (2023).






\bibitem{QFI1}
S. L. Braunstein and C. M. Caves, Statistical distance and the geometry of quantum states, 
Phys. Rev. Lett. \textbf{72}, 3439 (1994).

\bibitem{QFI2}
M. G. A. Paris, Quantum estimation for quantum technology, 
Int. J. Quant. Inf. \textbf{7}, 125 (2009).

\bibitem{nomono1}
J. Feng and J.-J. Zhang, Quantum Fisher information as a probe for Unruh thermality, Phys. Lett. B \textbf{827}, 136992 (2022).





\bibitem{Disc1}
W. H. Zurek, Pointer basis of the quantum apparatus, into what mixture does the wave packet collapse? 
Phys. Rev. D \textbf{24}, 1516 (1981).



\bibitem{Disc2}
J. P. Santos, L. C. C\'eleri, G.T. Landi and M. Paternostro, The role of quantum coherence in non-equilibrium entropy production, npj Quant. Inf. \textbf{5}, 23 (2019).









\end{thebibliography}
\end{document}